\documentclass[conference,table]{IEEEtran}
\usepackage[letterspace=-90]{microtype}
\usepackage{amsmath,amssymb,amsfonts,enumerate,multirow,xfrac,url,bm}
\usepackage{algorithmic,lipsum}
\usepackage{graphicx}
\usepackage{textcomp}
\usepackage{xcolor}
\usepackage{upgreek}
\usepackage{manfnt,yfonts}
\usepackage{algorithm,algorithmic}
\usepackage[caption=false, font=footnotesize]{subfig}
\graphicspath{{./images/}}

\usepackage{definice}
\usepackage{mathrsfs}

\usepackage{xcolor}

\usepackage{float}

\newcommand{\imageHeight}{\textls{\texttt{image\_\,height}}}
\newcommand{\imageWidth}{\textls{\texttt{image\_\,width}}}

\def\e{\mathrm{e}}
\newcommand{\T}{^{\top}}
\newcommand{\Expect}[1]{\mathrm{E}\!\left(\, #1\, \right)}
\newcommand{\cov}[1]{\text{cov}\!\left(\, #1\, \right)}
\newcommand{\var}[1]{\text{var}\!\left(\, #1\, \right)}

\newcommand{\ms}{\hspace{0.17cm}} 
\newcommand{\px}{\texttt{|\hspace{-0.03cm}p\hspace{-0.03cm}x\hspace{-0.03cm}|}}

\newcommand{\meanScalar}[1]{m_{#1}}

\newcommand{\xTwoDee}{\mathsf{x}}

\newcommand{\yTwoDee}{\mathsf{y}}

\newcommand{\widthTwoDee}{\upomega}

\newcommand{\heightTwoDee}{\mathsf{h}}

\newcommand{\xThreeDee}{x}
\newcommand{\xVelocityThreeDee}{\dot{x}}
\newcommand{\yThreeDee}{y}
\newcommand{\yVelocityThreeDee}{\dot{y}}
\newcommand{\zThreeDee}{z}
\newcommand{\zVelocityThreeDee}{\dot{z}}
\newcommand{\widthThreeDee}{\omega}
\newcommand{\heightThreeDee}{h}

\newcommand{\StateSet}{X}
\newcommand{\SurvRFS}{S}
\newcommand{\BornRFS}{B}
\newcommand{\ObjectGenRFS}{D}
\newcommand{\ClutterRFS}{C}

\newcommand{\blkdiag}[1]{\text{blkdiag} \hspace{-0.05cm} \left(\, {#1} \, \right)}

\newcommand{\Markov}{M}
\newcommand{\likelihood}{L}

\begin{document}

\IEEEoverridecommandlockouts
\title{
    Model-based Multi-object Visual Tracking: \\ Identification and Standard Model Limitations
	\thanks{
        This research was funded by the project SGS-2025-020.
	}
}

\author{
 \IEEEauthorblockN{Jan Krejčí\IEEEauthorrefmark{1}, Oliver Kost\IEEEauthorrefmark{1}, Yuxuan Xia\IEEEauthorrefmark{2}, Lennart Svensson\IEEEauthorrefmark{3}, Ondřej Straka\IEEEauthorrefmark{1}}
 \IEEEauthorblockA{
\IEEEauthorrefmark{1} \textit{Department of Cybernetics, University of West Bohemia}, Pilsen, Czech Republic
 }
 \IEEEauthorblockA{
 \IEEEauthorrefmark{2}\textit{Department of Automation, Shanghai Jiaotong University}, Shanghai, China
 }
 \IEEEauthorblockA{
 \IEEEauthorrefmark{3}\textit{Signal Processing Group, Chalmers University of Technology}, Göteborg, Sweden
 }
 \IEEEauthorblockA{
 Email: \{ jkrejci, kost, straka30 \}@kky.zcu.cz, yuxuan.xia@sjtu.edu.cn, lennart.svensson@chalmers.se
 }
 }

\maketitle

\begin{abstract}
    This paper uses multi-object tracking methods known from the radar tracking community to address the problem of pedestrian tracking using 2D bounding box detections. 
    The standard point-object (SPO) model is adopted, and the posterior density is computed using the Poisson multi-Bernoulli mixture (PMBM) filter.
    The selection of the model parameters rooted in continuous time is discussed, including the birth and survival probabilities.
    Some parameters are selected from the first principles, while others are identified from the data, which is, in this case, the publicly available MOT-17 dataset. 
    Although the resulting PMBM algorithm yields promising results, a mismatch between the SPO model and the data is revealed.
    The model-based approach assumes that modifying the problematic components causing the SPO model-data mismatch will lead to better model-based algorithms in future developments.
\end{abstract}


\section{Introduction}\label{sec:introduction}
Visual multi-object tracking is a key technology for public safety monitoring, autonomous driving, and many others~\cite{KITTI-dataset:2012,Dendorfer:MOTChallenge:2021,VisDrone:2021}.
This paper examines the use of a single monocular camera to track objects represented by two-dimensional bounding boxes (BBs), specifically pedestrians.
For such a setting, many methods have been developed in the field of computer vision~(CV) over the past decade that yield exceptional results.
This is documented, e.g., in the website\footnote{
    \underline{https://motchallenge.net/data/MOT17/}
} of the MOTChallenge visual tracking benchmark~\cite{Dendorfer:MOTChallenge:2021}.

The CV-based methods usually exploit the image information via \emph{(i)} a visual detection network (VDN) that outputs BB detections for each frame separately
and \emph{(ii)} detection scores; custom appearance, \mbox{re-ID}, or other features.
The \mbox{CV-based} methods, however, fuse the information in a heuristic manner\footnote{
    Although the algorithms commonly claim they use the Kalman Filter, it is strictly speaking not the case~\cite{KrKoStDu:2024:IEEE}.
    The estimation error covariance matrices are usually not presented to the user, and they fail to reflect the true uncertainty of the estimates~\cite{KrKoStDu:2024:Simple:IEEE}.
    Note that the performance evaluation used in, e.g., the MOTChallenge~\cite{Dendorfer:MOTChallenge:2021} does not take estimation uncertainty into account and was argued to be questionable for several reasons, see~\cite{Nguyen:Visual-metrics-trustworthy:2023,KrKoStXiSvFe:TGOSPA-params-visual:ArXiv}.
} and are often application-specific.
They have parameters that mostly lack a sound interpretation and thus also lack sound guidelines for their proper selection.
Their integration with other sensors can rely solely on heuristic workarounds, and components of the algorithms may be hard to replace. 
As a result, they lack interpretability and modularity, which leads to poor transferability to other problems.
This raises concerns about the use of CV-based methods in the previously mentioned safety-critical applications.

In the radar tracking community, model-based methods have been introduced to develop physically sound approximate algorithms for low-level information fusion~\cite{Mahler-Book:2007}.
This approach potentially applies to a wide range of problems and offers high modularity (e.g., for using several sensors of different nature jointly) under the Bayesian paradigm.
The \emph{probability hypothesis density} (PHD) filter~\cite{Mahler-PHD:2003},
the \emph{labeled multi-Bernoulli} (LMB) filter~\cite{LMB:2014}
or the \emph{Poisson multi-Bernoulli mixture} (PMBM) filter~\cite{PMBM:2018} are classic examples.
These algorithms are designed to use a general multi-object model, commonly known as the \emph{standard point-object} (SPO) model, whose parameters are well-defined and have a clear interpretation.
These algorithms provide a strong foundation for various tracking problems beyond radar tracking.

The model-based methods have been adopted in the CV context multiple times.
To name a few, the Gaussian mixture (GM) PHD filter was employed in~\cite{GMPHDOGM17:2019} and~\cite{Baisa:GMPHD-Occlu:2021}, the LMB filter was adopted in~\cite{Linh:VisualOccluLMB:2024} and the PMBM filter was used in~\cite{Mono-Camera3D_PMBM:2018}.
To improve performance and stay competitive in tracking benchmarks, designers adjust the algorithms instead of altering the underlying models.
This results in ad-hoc parameter tuning and the addition of heuristic components, such as occlusion handling strategies and re-ID modules, on top of model-based methods. Consequently, the algorithms suffer from the previously mentioned drawbacks of the CV-based methods.

This paper takes a first step towards developing high-performing visual tracking methods using model-based strategies, by investigating various assumptions and parameter choices. 
All parameters of the SPO model are discussed based on~\cite{KrKoStDu:2024:IEEE,KrKoSt:2023_FUSION,ContinuousDiscrete-MTT:2020}, which emphasize first principles.
Specific values of the parameters are given for the context of pedestrian tracking.
Similarly to the well-known CV-based \emph{Simple online and real-time tracking} (SORT)~\cite{SORT:2016}, only VDN bounding box outputs at each image frame are considered as the inputs to the algorithm.
In the case of Poisson birth, the corresponding optimal model-based filter is the PMBM filter, which is implemented using standard \emph{multi-hypothesis tracking} (MHT) approximations and shown to yield promising results.
In this paper, the suitability of the SPO model for the problem at hand is assessed, and several problematic components of the model are identified.
Thanks to the model-based approach, the problematic components directly indicate how the SPO model could be \emph{adjusted} to better align with reality yielding more reliable algorithms that ultimately lead to improved results.

This paper is organized as follows.
Section~\ref{sec:MTT-general} reviews the model-based approach and introduces the SPO model.
For the context of visual tracking, the SPO model is specified in Section \ref{sec:SPO-Visual-tracking-model}.
The corresponding model-based filter is tested on real data in Section~\ref{sec:results} together with a suitability assessment of the underlying SPO model.
The paper concludes in Section~\ref{sec:conclusion}.

\section{Multi-Object Modeling and Tracking}\label{sec:MTT-general}
This section briefly reviews the \emph{random finite set} (RFS) approach to multi-object modeling and Bayes filtering~\cite{Mahler-Book:2007}.
The SPO model is given in general terms, and the corresponding model-based filter, i.e., the PMBM filter, is discussed. 

\subsection{Model-based Approach to Multi-object Tracking}
Throughout the paper, the symbol $\StateSet_k {=} \{ \mathbf{x}_k^1,\dots,\mathbf{x}_k^{n_k} \}$ denotes the set of object states at time step $k{=}0,\dots,K$, with $K {\in} \mathbb{N}$ being the final time step.
Briefly speaking, both the cardinality $n_k {=} |\StateSet_k|$ and the individual states $\mathbf{x}_k^i$, $i{=}1,\dots,n_k$ are modeled as random, see~\cite{Mahler-Book:2007}.
Similarly, measurements received at each time step are modeled as RFSs $Z_k {=} \{ \mathbf{z}_k^1, \dots, \mathbf{z}_k^{m_k}\}$, $k{=}0,\dots, K$, where $\mathbf{z}_k^j$, $j{=}1,\dots,m_k$ will be taken as the VDN BBs.
For simplicity, random entities and their realizations are denoted by the same symbol. 

For $k{=}0,\dots,K$, the already-received measurement sets $Z^k {\triangleq} (Z_0,\dots,Z_k)$ are used to yield the posterior density $p(X_k|Z^k)$ of $X_k$ via the recursive Bayes filter~\cite{Mahler-Book:2007}
\begin{subequations}\label{eqs:MTT-Bayes-filter}
\begin{align}
    p \big(X_k | Z^k \big) &\propto p(Z_k | X_k) \cdot p(X_k | Z^{k-1}), \\
    p(X_{k} | Z^{k-1}) &= \textstyle\int p(X_{k}|X_{k-1}) p(X_{k-1}|Z^{k-1}) \delta X_{k-1},
\end{align}
\end{subequations}
where $p(X_0|Z^{-1})\triangleq p(X_0)$ is the initial density, $p (X_{k+1} | Z^{k})$ is the predictive density and the densities $p(Z_k | X_k)$ and $p(X_{k+1}|X_{k})$ encapsulate the measurement and motion models such as the SPO model, respectively.
The symbol $\propto$ denotes equality up to a normalizing constant.
The integral of a function $f(X)$ with finite-set inputs is taken to be the \emph{set integral}~\cite{Mahler-Book:2007}
\begin{align}    
    \textstyle \int f(X) \delta X \triangleq \sum_{n=0}^{+\infty} \tfrac{1}{n!} \int f (\{ \mathbf{x}^1,\dots,\mathbf{x}^n \}) \mathrm{d}\mathbf{x}^1\cdots\mathrm{d}\mathbf{x}^n,
\end{align}
and the densities are such that $\int p(X)\delta X = 1$.
The following example densities are essential for the construction of both the SPO model and the corresponding Bayes filter~\eqref{eqs:MTT-Bayes-filter} itself.

\subsubsection{Bernoulli RFS}
A Bernoulli RFS is either empty or a singleton.
The corresponding density function is
\begin{align}
    p_{  \mathrm{Ber} }(\StateSet) = \begin{cases}
        1-r & \text{if } \StateSet = \emptyset, \\[-0.1cm]
        r \cdot p_{\mathrm{sp}}(\mathbf{x}) & \text{if } \StateSet = \{\mathbf{x}\}, \\[-0.1cm]
        0 & \text{otherwise},
    \end{cases}
\end{align}
where $r\in[0,1]$ is the existence probability and $p_{\mathrm{sp}}(\mathbf{x})$ is the spatial probability density of $\mathbf{x}$ conditioned on its existence.

\subsubsection{Multi-Bernoulli RFS}
A multi-Bernoulli (MB) RFS is a union of multiple independent Bernoulli RFSs.
Denoting the density of the $\ell$-th Bernoulli RFS with $p_{\mathrm{Ber}}^\ell (X)$, $\ell{=}1,\dots,N$, the density of their union -- of the MB RFS -- is given by
\begin{align}
    p_{ \mathrm{MB} }(\StateSet) = \textstyle \sum_{ \uplus_{\ell=1}^{N} Y^\ell = \StateSet} \, \prod_{\ell=1}^N p_{  \mathrm{Ber} }^\ell ( Y^\ell ) \, ,
\end{align}
where the summation involving the symbol $\uplus$ is taken over all mutually \emph{disjoint} sets $Y^1,\dots,Y^N$ whose union is equal to $\StateSet$.

\subsubsection{Poisson RFS}
A Poisson RFS describes an independent and identically distributed set of points (states) whose number is Poisson distributed.
The corresponding density function is
\begin{align}
    p_{ \mathrm{Pois} } (\StateSet) = \textstyle \e^{-\rho } \prod_{\mathbf{x} \in \StateSet} \, \rho \cdot p_{\mathrm{sp}}(\mathbf{x}) \, ,
\end{align}
where $\rho \geq 0$ is the expected number of points and $p_{\mathrm{sp}}(\mathbf{x})$ the spatial probability density. 

\subsubsection{Poisson Multi-Bernoulli RFS}
The union $X=X^1\uplus X^2$ of independent Poisson-distributed RFS $X^1$ and MB-distributed RFS $X^2$ is the Poisson multi-Bernoulli (PMB) RFS.
The corresponding density function is given by
\begin{align}
    p_{\mathrm{PMB}}(X) = \textstyle \sum_{ Y^1 \uplus Y^2 = \StateSet} \, p_{\mathrm{Pois}} \big(Y^1\big) \cdot p_{  \mathrm{MB} } \big( Y^2 \big) \, . \label{eq:PMB-density-function}
\end{align}

\subsection{SPO Measurement Model}
In the SPO model, each object $\mathbf{x}_k {\in} \StateSet_{k}$ is either detected with the state-dependent \emph{probability of detection} $P_\ObjectGenRFS(\mathbf{x}_k)$ and independently generates a measurement according to the single-object likelihood function $\likelihood(\mathbf{z}_k|\mathbf{x}_k)$, or it is \emph{undetected} with the probability $1 {-} P_\ObjectGenRFS(\mathbf{x}_k)$.
Moreover, clutter measurements, independent of object-originating measurements, may appear in the measurement set, being described with the \emph{clutter} RFS~$\ClutterRFS_k$.
The set of all measurements at time step $k$ is thus~\cite[pp.~411]{Mahler-Book:2007},
\begin{align}\label{eq:MTT_MeasurementRFS_Standard-point-object}
	Z_k =
            \Big( \, \textstyle \bigcup_{\mathbf{x}_k\in\StateSet_k} \ObjectGenRFS(\mathbf{x}_k) \, \Big) 
            \cup
            \ClutterRFS_k
\end{align}
where $\ObjectGenRFS(\mathbf{x}_{k})$ is a conditional Bernoulli RFS with the PDF
\begin{align}\label{eq:standard-point-object_measurement-object-gen_Bernoulli}
	p
    (D|\mathbf{x}_{k}) = \begin{cases}
		1-P_\ObjectGenRFS(\mathbf{x}_{k}), & \text{if } D = \emptyset, \\
		P_\ObjectGenRFS(\mathbf{x}_{k})\likelihood(\mathbf{z}_k | \mathbf{x}_{k} ), & \text{if } D = \{\mathbf{z}_k\}, \\
		0, & \text{otherwise}.
	\end{cases}
\end{align}
The clutter RFS $\ClutterRFS_k$ is modeled as a Poisson RFS with the PDF
\begin{align}
    p(\ClutterRFS_k) = \textstyle \e^{-\lambda} \prod_{\mathbf{z}_k \in \ClutterRFS_k} \lambda c(\mathbf{z}_k),
    \label{eq:clutter-PPP}
\end{align}
where $\lambda \geq 0$ is the expected number of clutter measurements and $c(\mathbf{z})$ is the clutter spatial PDF.

\subsection{SPO Motion Model}
Each object $\mathbf{x}_{k-1} {\in} \StateSet_{k-1}$ from time step $k{-}1$ either survives to time step $k$ with the state-dependent \emph{probability of survival} $P_\SurvRFS(\mathbf{x}_k)$ and moves independently according to the single-object Markov transition function $\Markov(\mathbf{x}_k|\mathbf{x}_{k-1})$, or it disappears with the probability $1 {-} P_\SurvRFS(\mathbf{x}_k)$.
Moreover, new objects, independent of the existing objects, may appear in the tracking area, described with the \emph{birth} RFS $\BornRFS_k$, leading to \cite[pp.~467]{Mahler-Book:2007}
\begin{align}\label{eq:MTT_MotionRFS_Standard-point-object}
	\StateSet_k = 
        \Big( \, \textstyle \bigcup_{\mathbf{x}_{k-1}\in\StateSet_{k-1}} \SurvRFS(\mathbf{x}_{k-1}) \, \Big) 
        \cup 
        \BornRFS_k
\end{align}
where $\SurvRFS(\mathbf{x}_{k-1})$ is a conditional Bernoulli RFS with the PDF
\begin{align}\label{eq:SurvivalRFS-PDF}
	p (S|\mathbf{x}_{k-1}) \! = \! \begin{cases}
		1-P_\SurvRFS(\mathbf{x}_{k-1}), & \text{if } S = \emptyset, \\
		P_\SurvRFS(\mathbf{x}_{k-1})\Markov(\mathbf{x}_k | \mathbf{x}_{k-1} ), & \text{if } S = \{\mathbf{x}_k\}, \\
		0, & \text{otherwise}.
	\end{cases}
\end{align}
The birth RFS $\BornRFS_k$, also known as \emph{birth} or \emph{arrival} process, can be either MB or Poisson.
The following notation is used.

\subsubsection{Multi-Bernoulli Birth}
Let $N_{k,\BornRFS}\in\mathbb{N}$ Bernoulli birth components be given at time step $k$, whose existence probabilities and spatial densities are denoted with $r_{k,\BornRFS}^{\ell}$ and $b_{k,\BornRFS}^{\ell}(\mathbf{x}_k)$, respectively,  for $\ell{=}1,\dots,N_{k,\BornRFS}$.

\subsubsection{Poisson Birth}
Without a lack of generality, let the Poisson birth be given by the best Poisson approximation of the above MB birth model in the sense of~\cite[p.~579]{Mahler-Book:2007}. 
That is, the expected number of newborn objects and the spatial distribution for the Poisson RFS are taken to be
\begin{subequations}\label{eqs:Poisson-birth:MB-best-fitting}
\begin{align}
    \beta_k &= \textstyle\sum_{\ell = 1}^{N_{k,\BornRFS}} r_{k,\BornRFS}^{\ell}, \label{eq:expected-number-of-newly-born-objects} \\
    b_k (\mathbf{x}_k) &= \textstyle\sum_{\ell = 1}^{N_{k,\BornRFS}} \tfrac{1}{\beta_k} {\cdot} r_{k,\BornRFS}^{\ell} \cdot b_{k,\BornRFS}^{\ell}(\mathbf{x}_k), \label{eq:spatial-density-of-newly-born-objects}
\end{align}
\end{subequations}
respectively.
Notice that the above Poisson birth is \emph{less}~informative compared to the MB birth, which is preferable if no prior knowledge regarding the birth is given.
With the Poisson birth and the individual motion model~\eqref{eq:SurvivalRFS-PDF}, the density $p(X_k|X_{k-1})$ resulting from~\eqref{eq:MTT_MotionRFS_Standard-point-object} has a form of a PMB density~\eqref{eq:PMB-density-function} as well.

\subsection{Bayes Filter Corresponding to the SPO Model}
The SPO model is adopted by a number of tracking solutions\footnote{
    Assuming the birth and the initial density are both MB (or mixtures of thereof), the corresponding model-based filter is the multi-Bernoulli mixture (MBM) filter~\cite{Mahler-integralTransform:2016,GM-MBM:2019}.
    So-called \emph{labels} can be used in the formulation, leading to LMB and \emph{generalized labeled multi-Bernoulli} (GLMB)~\cite{VoVo-dGLMB:2014} filters.
}.
In this paper, the birth and the initial distribution are assumed to be Poisson.
With these modeling assumptions,
the corresponding model-based filter is given exactly by the PMBM filter~\cite{PMBM:2018}. 
That is, the posterior and predictive multi-object densities for all $k$ are in the form of a \emph{mixture} of PMB densities with the form~\eqref{eq:PMB-density-function}.
Note that the best Poisson approximation to the PMBM posterior can be computed in closed form, which leads to the PHD filter~\cite[Section~16.3]{Mahler-Book:2007}, see also~\cite{Williams-U-PMBM:2012}.

In the PMBM filter, individual objects are represented with the Bernoulli components, whose initialization is measurement-driven.
Each term of the mixture represents the posterior density of a set of detected objects conditioned on a particular data association hypothesis, while the Poisson part represents objects that have never been detected. 
The recursion is omitted in this paper and can be found in~\cite{PMBM:2018}.\\

\begin{table}
	\centering
	\caption{
        List of parameters of the standard point-object model.
        }\label{tab:parameters_standard-point-object}
	\begin{tabular}{ccl}
		counterpart & symbol & meaning \\ \hline
		measurement 
		& $\likelihood(\mathbf{z}_k | \mathbf{x}_k)$ & single-object likelihood function \\
        & $P_\ObjectGenRFS(\mathbf{x}_k)$ & probability of detection \\
		& $\lambda$ & clutter: expected number \\
		& $c(\mathbf{z}_k)$ & clutter: spatial density \\ \hline
		motion 
		& $\Markov(\mathbf{x}_k|\mathbf{x}_{k-1})$ & single-object transition function \\
        & $P_\SurvRFS(\mathbf{x}_{k-1})$ & probability of survival \\
        & $\beta_k$ & birth: expected number \\
        & $b_k(\mathbf{x}_k)$ & birth: spatial density
	\end{tabular}
    \vspace{-2mm}
\end{table}

To apply the model-based filter in practice, the parameters of the model must be given.
In general, the parameters are mostly functions and are listed in Table~\ref{tab:parameters_standard-point-object}.
Even though the PMBM filter in such general settings can be implemented with, e.g., sequential Monte Carlo methods \cite{SMCinPractice:2001}, practical implementations often assume a linear-Gaussian model of individual targets and constant probabilities of detection $P_\ObjectGenRFS$ and survival $P_\SurvRFS$.
Note that clutter spatial distribution $c(\mathbf{z})$~\eqref{eq:clutter-PPP} can be arbitrary for such implementations. 
Thanks to the intuitive interpretation, users can typically select parameters manually for specific applications.
In the following, the parameters are specified for visual tracking based on~\cite{KrKoStDu:2024:IEEE,KrKoSt:2023_FUSION,ContinuousDiscrete-MTT:2020}, leading to an \emph{approximate} Gaussian mixture implementation of the PMBM filter~\cite[Sec.~V]{PMBM:2018}, further abbreviated as GM-PMBM.

\section{Parameters Suitable for Visual Tracking}\label{sec:SPO-Visual-tracking-model}
In this section, the single-object state-space model introduced in~\cite{KrKoStDu:2024:IEEE} and based on first principles is accompanied by the SPO (i.e., multi-object) model parameters.
In particular, measurement model parameters are discussed based on data~\cite{KrKoSt:2023_FUSION}, and physics-based modeling of the survival and birth parameters are based on~\cite{ContinuousDiscrete-MTT:2020}.
The parameter values are exemplified for using the Faster R-CNN detector applied to the MOT-17 dataset (training subset)~\cite{MOT-16:2016,MOT17-webpage:2023}.
For simplicity, the camera is assumed to be static, and only BBs are taken into account, i.e., not detection scores or features.

\begin{figure}[h] 
	\centering
        \vspace{-2mm} 
	\includegraphics[width=0.8\linewidth]{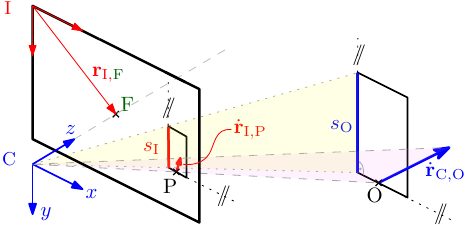}
        \vspace{-2mm} 
	\caption{
    Illustration of the pinhole camera geometry.
    A height $s_{\mathrm{O}}$ in meters and velocity $\dot{\mathbf{r}}_{\mathrm{C,O}}$ in m$\cdot$s$^{-1}$ described in the camera coordinates $\mathrm{C}$ are \emph{projected} to the corresponding variables $s_{\mathrm{I}}$ in pixels and $\dot{\mathbf{r}}_{\mathrm{I,P}}$ in px$\cdot$s$^{-1}$ described in the image coordinates $\mathrm{I}$.
    The reference point of the object $\mathrm{O}$ is projected to the point $\mathrm{P}$.
    The vector $\mathbf{r}_{\mathrm{I,F}}$ in px points from $\mathrm{I}$ to the \emph{focal} point $\mathrm{F}$.
    }\label{fig:camera_geometry_2}
        \vspace{-2mm} 
\end{figure}

Given the pinhole camera model, a pedestrian is modeled in~\cite{KrKoStDu:2024:IEEE} with a \emph{planar BB in 3D}.
The \emph{state} at time-step $k$ is 
\begin{align}
\nonumber\\[-6mm]
    \mathbf{x}_k = [ \xThreeDee_k\ \xVelocityThreeDee_k\ \yThreeDee_k\ \yVelocityThreeDee_k\ \zThreeDee_k\ \zVelocityThreeDee_k\ \widthThreeDee_k\ \heightThreeDee_k ]\T \in \mathbb{R}^8, \label{eq:3Dstate}
\\[-6mm]\nonumber
\end{align}
where the position $\xThreeDee_k$, $\yThreeDee_k$, $\zThreeDee_k$ of the bottom center of the box and its width $\widthThreeDee_k$ and height $\heightThreeDee_k$ are in meters, while the velocities $\xVelocityThreeDee_k$, $\yVelocityThreeDee_k$, $\zVelocityThreeDee_k$ are in meters per second.
An example of such a 3D BB state is given in Fig.~\ref{fig:camera_geometry_2}, in which case the position $[\xThreeDee_k\ \yThreeDee_k\ \zThreeDee_k]\T$ would be the vector pointing from $\mathrm{C}$ to $\mathrm{O}$, the velocity is $\dot{\mathbf{r}}_{\mathrm{C,O}}=[\xVelocityThreeDee_k\ \yVelocityThreeDee_k\ \zVelocityThreeDee_k]\T$, the height is $s_{\mathrm{O}}=\heightThreeDee_k$.

\subsection{Measurement Model Parameters for Visual Tracking}
Each output of a VDN for a frame $k$ is a detection $\mathbf{z}_k$ forming a BB with its position and extent modeled as a vector
\begin{align}
    \mathbf{z}_k = [\xTwoDee_k\ \yTwoDee_k\ \widthTwoDee_k\ \heightTwoDee_k]\T \in\mathbb{R}^4, \label{eq:annotation}
\end{align}
where $\xTwoDee_k$ and $\yTwoDee_k$ are the position coordinates of the bottom center point of the BB, while $\widthTwoDee_k$ and $\heightTwoDee_k$ are its width and height, respectively, all in pixels.
Given the state $\mathbf{x}_k$~\eqref{eq:3Dstate}, the measurement equation is~\cite{KrKoStDu:2024:IEEE}
\begin{align}
	\mathbf{z}_k = \mathbf{h}(\mathbf{x}_k)  + \mathbf{v}_k, \label{eq:measurement-equation}
\end{align}
where $\mathbf{v}_k$ is the measurement noise.
The nonlinear measurement function representing the perspective projection is given by
\begin{align}
\nonumber\\[-7mm]
	\mathbf{h}(\mathbf{x}_k) =
        \frac{f}{ \px \cdot [\mathbf{x}_k]_{5} } 
        \left[\begin{smallmatrix}
			[\mathbf{x}_k]_{1} \\[0.01cm]
			[\mathbf{x}_k]_{3} \\[0.01cm]
			[\mathbf{x}_k]_{7} \\[0.01cm]
			[\mathbf{x}_k]_{8}
        \end{smallmatrix}\right]
		+
        \left[\begin{smallmatrix}
			[\mathbf{r}_{\mathrm{I | F}}]_{1} \\[0.01cm]
            [\mathbf{r}_{\mathrm{I | F}}]_{2} \\[0.07cm]
			0 \\[0.07cm]
			0
        \end{smallmatrix}\right]
        , \label{eq:T3d2d_state}
        \\[-6mm]\nonumber
\end{align}
where $[\, \cdot\, ]_{\ell}$ denotes the $\ell$-th element of the  vector, see Fig.~\ref{fig:camera_geometry_2}.
Pixels are assumed to be rectangular with side length $\px$ in meters, and $f$ is the focal length in meters.
If unknown, hand-selected values such as $\px {=} 10^{-6}~\mathrm{m}$ and $f {=} 10^{-3}~\mathrm{m}$ could be used~\cite{KrKoSt:2023_FUSION}.
For the Faster R-CNN (FRCNN) detector applied to the MOT-17 dataset~\cite{MOT-16:2016,MOT17-webpage:2023}, the measurement noise covariance matrix
\begin{align}
    \mathbf{R} = \cov{ \mathbf{v}_k } \!= \gamma^2
     \! \cdot \! 10^{-5} \! \left[\begin{smallmatrix}
    2.029 &\ 0.223 &\ 0.073 &\ 0.248 \\
    0.223 &\ 3.051 &\ 2.549 &\ 0.285 \\
    0.073 &\ 2.549 &\ 4.880 &\ 0.179 \\
    0.248 &\ 0.285 &\ 0.179 &\ 2.032 
    \end{smallmatrix}\right] ,
    \label{eq:state-noise-covariance-matrix:identification}
\end{align}
with $\gamma=\min( \imageWidth, \imageHeight )$,
was estimated using the method\footnote{ \label{footnote:estimation-tlwh-visibility}
    In~\cite{KrKoSt:2023_FUSION}, the matrix $\mathbf{R}$ is denoted as $\mathbf{R}_{\mathrm{U}}$ and it was estimated for a different measurement representation:
    the measurement in~\cite{KrKoSt:2023_FUSION} included the top-left corner of the bounding box instead of the bottom center, which is used in this paper, and in~\cite{KrKoStDu:2024:IEEE,KrKoStDu:2024:Simple:IEEE}.
    It was thus used erroneously in~\cite{KrKoStDu:2024:IEEE,KrKoStDu:2024:Simple:IEEE}.
    
    Further, note that no minimum visibility was used to drive the estimated $\mathbf{R}$~\eqref{eq:state-noise-covariance-matrix:identification} unlike in~\cite{KrKoSt:2023_FUSION} where a minimum visibility of $0.1$ was used.
    The same is true for the estimated constant probability of detection $P_{\ObjectGenRFS}$ and the expected number of clutter measurements $\lambda$.
}~\cite{KrKoSt:2023_FUSION}.

\subsubsection{Single-object Likelihood Function}\label{sbusbusec:likelihood-function}
A Gaussian distribution is known to impose the minimal prior structural constraints when only the first two moments are known~\cite[pp.~253-255]{Cover:ElementsOfInformationTheory:2006}.
The assumption $\mathbf{v}_k \sim \mathcal{N}(\mathbf{0}, \mathbf{R}) $ thus adds as little information as possible into the construction of the likelihood, which becomes
\begin{align}
    \likelihood(\mathbf{z}_k | \mathbf{x}_k) = \mathcal{N}\big( \mathbf{z}_k;\ \mathbf{h}(\mathbf{x}_k),\, \mathbf{R} \big) \, . \label{eq:single-object-likelihood-gaussian}
\end{align}
Since the measurement function $\mathbf{h}(\mathbf{x}_k)$ is nonlinear, the model-based filter does not generally admit a GM implementation. 
This inconvenience will be solved by using the \emph{approximate} unscented Kalman filter (UKF) as suggested in~\cite{KrKoStDu:2024:IEEE}.

\subsubsection{Probability of Detection}
To adopt the estimation method from~\cite{KrKoSt:2023_FUSION} directly, the probability of detection is assumed constant.
For the FRCNN detector, see footnote~\ref{footnote:estimation-tlwh-visibility}, we have
\begin{align}
    P_\ObjectGenRFS(\mathbf{x}_k) = P_\ObjectGenRFS = 0.529 \, . \label{eq:Pd-estimates}
\end{align}

\subsubsection{Clutter Model Parameters}
The estimated expected number of clutter measurements for the FRCNN detector using the method from~\cite{KrKoSt:2023_FUSION}, see footnote~\ref{footnote:estimation-tlwh-visibility}, is
\begin{align}
    \lambda = 1.552 \, . \label{eq:estimated-lambda-clutter}
\end{align}
The clutter spatial density is commonly modeled to be uniform over the tracking area.
Inspired by~\cite{KrKoSt:2023_FUSION}, we cover the possibility that BBs may be partially outside the image by using
\begin{align}
    c(\mathbf{z}_k) =\
    & p_{\mathrm{U}} \big( \xTwoDee_k;\ -\tfrac{1}{4}\cdot \imageWidth{}, \, \tfrac{5}{4}\cdot \imageWidth{} \big) \notag\\[-0.05cm]
    & \cdot p_{\mathrm{U}} \big( \yTwoDee_k;\ 0, \, \tfrac{3}{2}\cdot \imageHeight{} \big) \notag\\[-0.05cm]
    & \cdot p_{\mathrm{U}} \big( \widthTwoDee_k;\ 0, \, \tfrac{1}{2}\cdot \imageWidth{} \big) \notag\\[-0.05cm]
    & \cdot p_{\mathrm{U}} \big( \heightTwoDee_k;\ 0, \, \tfrac{4}{3}\cdot \imageHeight{} \big) , \label{eq:clutter-spatial}
\end{align}
where $p_{\mathrm{U}}( \,\cdot\, ; a, b)$ denotes the PDF of a uniformly distributed random variable on the interval $[a,b]$.

\subsection{Motion Model Parameters for Visual Tracking}
According to~\cite{KrKoStDu:2024:IEEE}, the dynamics of the state $\mathbf{x}_k$~\eqref{eq:3Dstate} for pedestrians is grounded in continuous-time modeling, yielding 
\begin{align}
	\mathbf{x}_{k+1} &=
	\mathbf{F} \mathbf{x}_{k}
	+
    \mathbf{m}
	+
	\mathbf{w}_{k} , \label{eq:3Dsystem:state} 
\end{align}
where the dynamic matrix and the additive term are
\begin{align}
    \mathbf{F} &= \blkdiag{ \left[\begin{smallmatrix} 1 & T \\ 0 & 1\end{smallmatrix}\right], \left[\begin{smallmatrix} 1 & T \\ 0 & 1\end{smallmatrix}\right], \left[\begin{smallmatrix} 1 & T \\ 0 & 1\end{smallmatrix}\right], \alpha_{\widthThreeDee}, \alpha_{\heightThreeDee} }, \label{eq:3Dsystem:dynamicMatrix}\\
    \mathbf{m} &= 
    \left[ \mathbf{0}_{1 \times 6} \ms
        (1\!-\!\alpha_{\widthThreeDee})\!\cdot\!\meanScalar{\widthThreeDee} \ms
        (1\!-\!\alpha_{\heightThreeDee})\!\cdot\!\meanScalar{\heightThreeDee} \right]\T,
\end{align}
respectively, where $\mathbf{0}_{m \times n}$ is a zero matrix of dimensions $m \times n$, $T$ is the sampling period, $\alpha_{\widthThreeDee} {=} e^{\sfrac{-\!T}{ \tau_{\widthThreeDee} }}$ and $\alpha_{\heightThreeDee} {=} e^{\sfrac{-\!T}{ \tau_{\heightThreeDee} }}$.
The height $\heightThreeDee_k$ of the planar 3D BB is modeled as that of a pedestrian, using the mean $\meanScalar{\heightThreeDee} {=} 1.65~\mathrm{m}$ and time-constant $\tau_{\heightThreeDee} {=} 4~\mathrm{s}$.
The width $\widthThreeDee_k$ of the planar 3D BB is modeled with the mean $\meanScalar{\widthThreeDee} {=} 0.85~\mathrm{m}$, and time-constant $\tau_{\widthThreeDee} {=} 0.4~\mathrm{s}$ to account for rough changes due to the hand swinging.
The state noise covariance matrix $\mathbf{Q} = \cov{\!\mathbf{w}_k\!}$ is
\begin{align}
	\mathbf{Q} \!=\! \blkdiag{ q_{\xVelocityThreeDee} \mathbf{T}, q_{\yVelocityThreeDee} \mathbf{T}, q_{\zVelocityThreeDee} \mathbf{T}, \sigma_{\widthThreeDee}^2(1\!-\!\alpha_{\widthThreeDee}^2), \sigma_{\heightThreeDee}^2(1\!-\!\alpha_{\heightThreeDee}^2) },
    \label{eq:3Dsystem:stateNoiseCov}
\end{align}
where the \emph{power spectral densities} $q_{\xVelocityThreeDee} {=} q_{\yVelocityThreeDee} {=} q_{\zVelocityThreeDee} {=} 1$~m$^2$s$^{-3}$ are set for pedestrians~\cite[pp.~418]{Groves-NavigationSystems:2013}, and the matrix
\begin{align}
    \mathbf{T} &= \left[\begin{smallmatrix} \tfrac{T^3}{3} & \tfrac{T^2}{2} \\ \tfrac{T^2}{2} & T\end{smallmatrix}\right] \, .
\end{align}
Finally, the standard deviations $\sigma_{\widthThreeDee} {=} \tfrac{0.45}{3}~\mathrm{m}$ and $\sigma_{\heightThreeDee} {=} \tfrac{0.3}{3}~\mathrm{m}$ of the width and height, respectively, are selected based on their assumed relation to the maximal deviations from the means, i.e., $\sigma_{\widthThreeDee}^{\mathrm{max}} {=} 0.45$~m and $\sigma_{\heightThreeDee}^{\mathrm{max}} {=} 0.3$~m, respectively.

\subsubsection{Single-object Transition Function}
Similarly to the construction of the likelihood function in Section~\ref{sbusbusec:likelihood-function}, 
the state noise $\mathbf{w}_k$ is assumed to be Gaussian. 
Therefore,
\begin{align}
    \Markov(\mathbf{x}_{k+1} | \mathbf{x}_k) = \mathcal{N}\big( \mathbf{x}_{k+1};\ \mathbf{F}\mathbf{x}_k {+} \mathbf{m},\, \mathbf{Q} \big) \, .
\end{align}

To comply with the continuous-time physical modeling, the probability of survival $P_\SurvRFS(\mathbf{x}_{k-1})$ and the expected number of newborn objects $\beta_k$~\eqref{eq:expected-number-of-newly-born-objects} are given by the M/M/$\infty$ queuing\footnote{
    The first letter denotes the type of birth (i.e., the arrival of new customers), whereas M stands for \emph{Markovian}, meaning that birth is described by the Poisson process in time.
    The second letter denotes the type of life span distribution, whereas M stands for \emph{Markovian} as well, meaning that life span is exponentially distributed.
    The last symbol is the number of servers, whereas $\infty$ means there is no queue at all, i.e., newborn objects do \emph{not} have to wait for existing objects to leave the system.
    For details, see, e.g.,~\cite[Sec.~16-3]{Papoulis:ProbabilityBook:2002}.
} system-based model of multi-object systems~\cite{ContinuousDiscrete-MTT:2020}.
Note that both parameters appear to be functions of $T$.

\subsubsection{Probability of Survival}
In the M/M/$\infty$ model, the life span of objects is assumed \emph{(i)} to be independent of the state $\mathbf{x}_{k-1}$ and \emph{(ii)} to follow exponential distribution with parameter $\mu {=} \tfrac{1}{L}$ where $L {>} 0$ in seconds is the mean life span of the objects.
It follows that the probability that an object stays in the system until the next time step, i.e., the probability of survival, is~\cite[Sec.~III.A]{ContinuousDiscrete-MTT:2020}
\begin{align}
    P_{\SurvRFS}(\mathbf{x}_{k-1}) = P_{\SurvRFS} = \e^{ -\sfrac{T}{L} } \, . 
    \label{eq:Ps-estimates}
\end{align}
Notice that the value $\e^{ -\sfrac{1}{L} }$ can be interpreted as the probability that an object survives over one second.

The mean life span $L$ can be hand-selected based on the captured view.
If ground-truth data is available, it may be estimated by averaging over the lifespans of individual objects, whereas the life span of an object is the number of frames it existed multiplied by $T$.
For pedestrians from the entire \mbox{MOT-17} dataset~\cite{MOT-16:2016,MOT17-webpage:2023} (training subset), the resulting average value is 
\begin{align}
    L = 7.481~\mathrm{s} \, . \label{eq:L-estimates}
\end{align}

\subsubsection{Birth Model -- Expected Number}\label{sec:birth:expected-number}
In the M/M/$\infty$ model, the birth is assumed to be a Poisson process.
That is, the number of newborn objects between any time instants $t_1$ and $t_2$, $t_2 {>} t_1$, is Poisson distributed with parameter $\eta {\cdot} (t_2 {-} t_1)$, where $\eta$ is the mean number of newborn objects per second.
It follows that the mean number of newborn objects between two consecutive time steps~\cite[Sec.~III.C]{ContinuousDiscrete-MTT:2020} is constant in time,
\begin{align}
    \beta_k = \beta = \eta \cdot L \cdot (1-P_\SurvRFS) \, .
\end{align}
Note that under the M/M/$\infty$ model, the number $L {\cdot} \eta$ is the mean number of objects $\Expect{ n_k }$ for $k {\rightarrow} \infty$.

The value $\eta$ can be hand-selected based on the captured view.
If ground-truth data is available, $\eta$ may be estimated as the total number of newly appearing objects divided by the dataset duration in seconds.
For pedestrians from the entire MOT-17 dataset~\cite{MOT-16:2016,MOT17-webpage:2023} (training subset), it is 
\begin{align}
    \eta = 1.925~\text{objects}/\mathrm{s} \, . \label{eq:eta-estimates}
\end{align}
Note that $\beta$ could be estimated directly from the data, but such an estimate requires data with the same
sampling period $T$.

\subsubsection{Birth Model -- Spatial Density}
The birth spatial density should account for location(s) where objects frequently appear for the first time, such as edges of the image frame or various entrances in 3D.
For both simplicity and generality, no particular locations are used in the following design. 
To permit an approximate \mbox{GM-PMBM} implementation, the spatial density~\eqref{eq:spatial-density-of-newly-born-objects} should be in the form of a GM\footnote{
    Without this constraint, the birth~\eqref{eqs:Poisson-birth:MB-best-fitting} could be simply designed with $N_{k,\BornRFS} {=} 1$ and $b_{k}(\mathbf{x}_k) {=} b_{k,\BornRFS}^{\ell=1}(\mathbf{x}_k)$ being uniform over the tracking area in 3D.
}, see~\cite[Sec.~V.B]{PMBM:2018}.
With this constraint, it is reasonable to design the birth~\eqref{eqs:Poisson-birth:MB-best-fitting} with $N_{k,\BornRFS} {=} 1$ and $b_{k}(\mathbf{x}_k) {=} b_{k,\BornRFS}^{\ell=1}(\mathbf{x}_k) {=} \mathcal{N}(\mathbf{x_k};\, \mathbf{m}_{k,\BornRFS}, \mathbf{P}_{k,\BornRFS})$, where $\mathbf{m}_{k,\BornRFS}$ is the scene center and $\mathbf{P}_{k,\BornRFS}$ is a covariance matrix designed to capture sufficient spatial uncertainty.

Since the measurement function $\mathbf{h}(\mathbf{x}_k)$~\eqref{eq:T3d2d_state} is nonlinear, the birth designed with a single Gaussian distribution may lead to poor approximation.
Moreover, considering the sigma-point-based UKF that will be used within the approximate \mbox{GM-PMBM} filter, birth covariance matrices should be designed with care.
To that end, the following simple design is proposed.

\begin{figure}
    \centering
    \includegraphics[width=0.49\linewidth,trim={1.0cm, 0.0cm, 0.5cm, 1.0cm},clip]{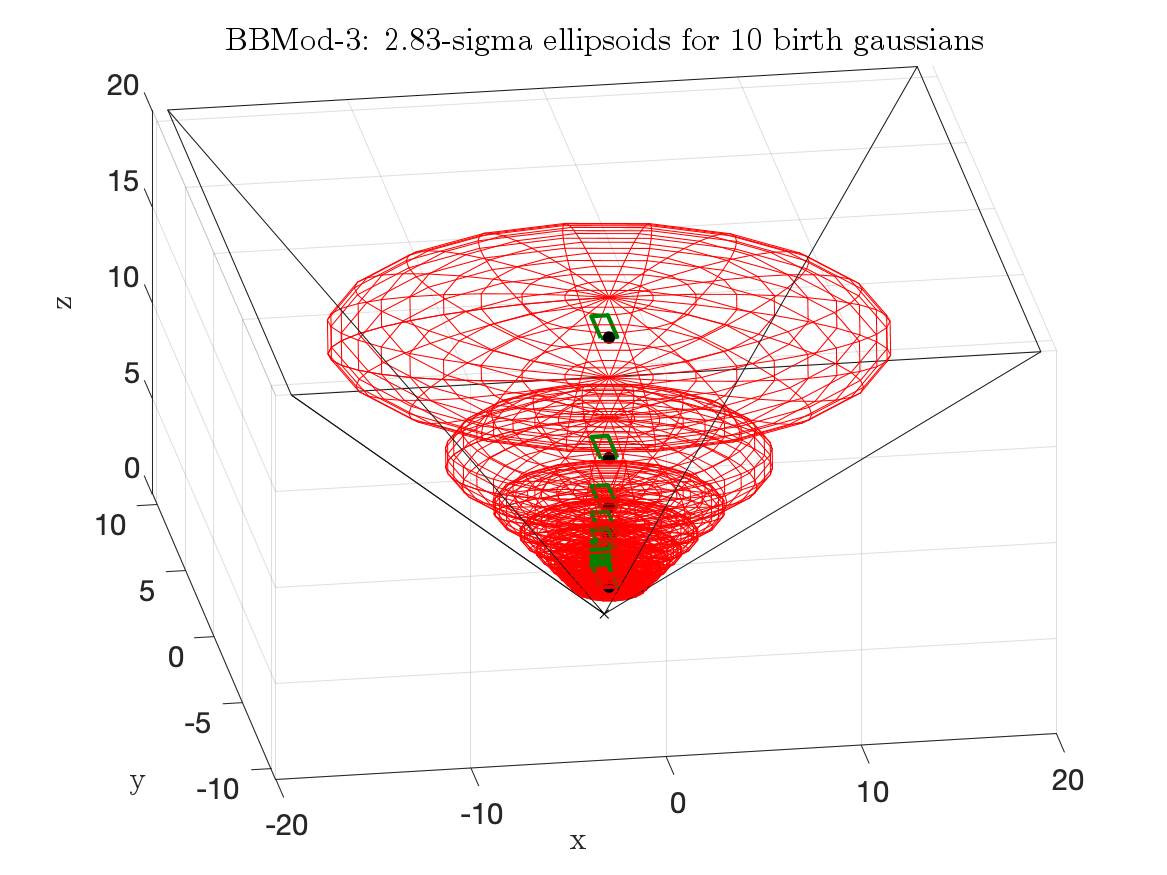}
    \includegraphics[width=0.49\linewidth,trim={2.6cm, 2.2cm, 2.6cm, 0.85cm},clip]{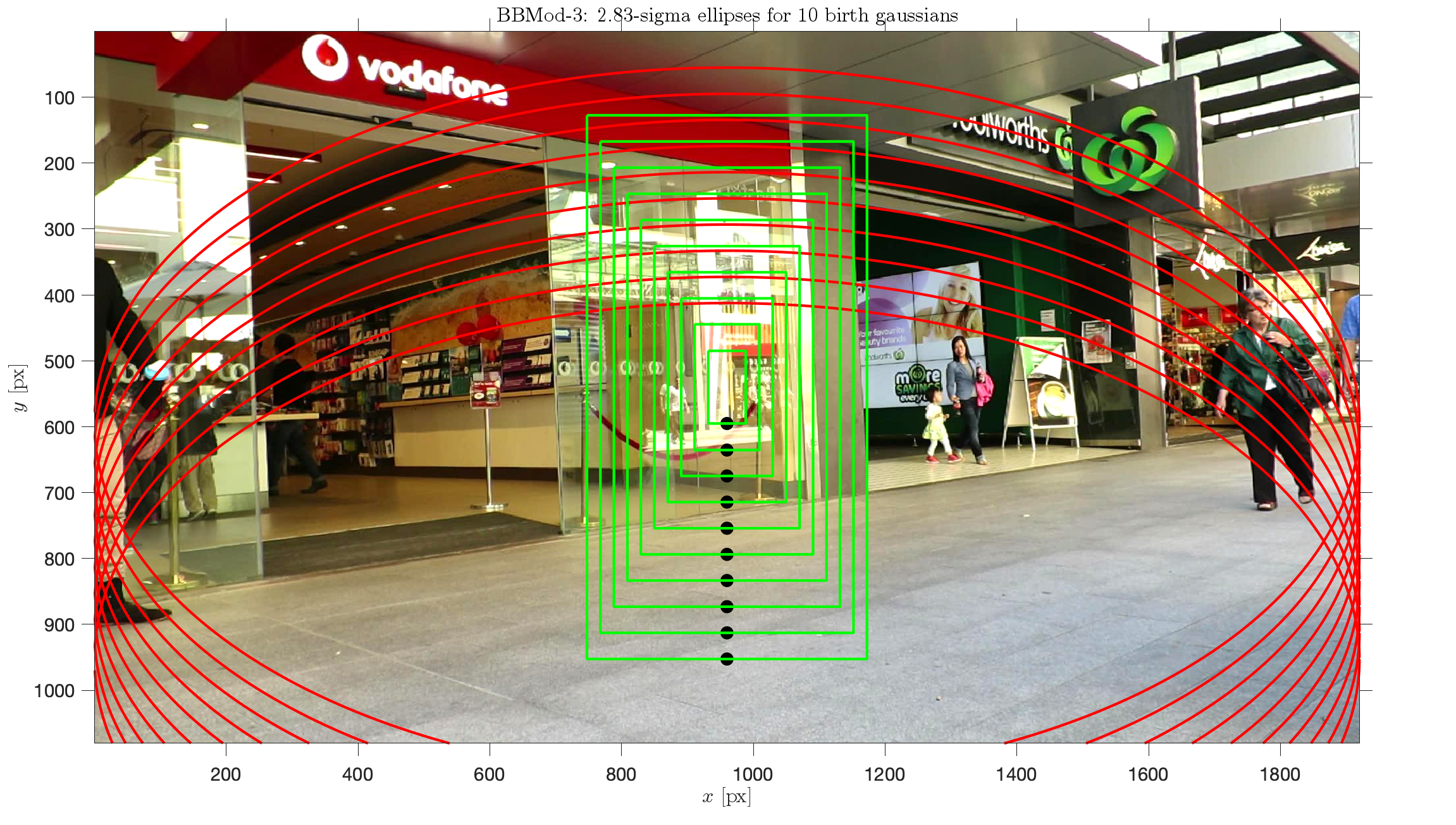}\\[-0.2cm]
    \caption{Design of $N_{k,\BornRFS} {=} 10$ components of the GM birth spatial density.
    The means $\mathbf{m}_{k,\BornRFS}^{\ell}$, $\ell {=} 1,\dots,10$ representing 3D BBs are plotted in green.
    The corresponding bottom center points are depicted as black points whose spatial diversities are illustrated with $\,\sqrt{8}$-sigma covariance ellipsoids depicted in red.
    Note that the sigma points used in the UKF approximation are drawn on the surface of the red ellipsoids.}
    \label{fig:birth-spatial}
    \vspace{-2mm}
\end{figure}

As the nonlinearity in $\mathbf{h}(\mathbf{x}_k)$~\eqref{eq:T3d2d_state} is due to the division by $[ \mathbf{x}_k ]_5 {=} \zThreeDee_k$, a given number $N_{k,\BornRFS} {\geq} 1$ of GM components can be designed to span the optical axis $\zThreeDee$ in 3D.
Since the smaller the $\zThreeDee_k$, the larger the nonlinearity, the means should be nonlinearly distributed along the $\zThreeDee$ axis.
As illustrated in Fig.~\ref{fig:birth-spatial}, the number of $N_{k,\BornRFS} {=} 10$ means $\mathbf{m}_{k,\BornRFS}^{\ell}$, $\ell {=} 1,\dots,10$ are laid from chosen minimum depth $\zThreeDee^{\min} {=} 2~\mathrm{m}$ to chosen maximum depth $\zThreeDee^{\max} {=} 15~\mathrm{m}$.
The mean velocities $ [\mathbf{m}_{k,\BornRFS}^{\ell}]_2$, $[\mathbf{m}_{k,\BornRFS}^{\ell}]_4$, $[\mathbf{m}_{k,\BornRFS}^{\ell}]_6$ are taken to be zero and the width $[\mathbf{m}_{k,\BornRFS}^{\ell}]_7 {=} \meanScalar{\widthThreeDee} {=} 1.65~\mathrm{m}$ and height $[\mathbf{m}_{k,\BornRFS}^{\ell}]_8 {=} \meanScalar{\heightThreeDee} {=} 0.85~\mathrm{m}$ are based on the motion model.
Furthermore, the covariance matrices $\mathbf{P}_{k,\BornRFS}^{\ell}$, $\ell {=} 1,\dots,10$ are designed as diagonal, based on the corresponding depth $ [\mathbf{m}_{k,\BornRFS}^{\ell}]_5 $ so that the UKF sigma-points are \emph{(i)} drawn on the image edges when considering $\xThreeDee$ and $\yThreeDee$ axes, and \emph{(ii)} do not overlap with neighboring sigma-points along the $\zThreeDee$ axis.
The variances of velocities are taken to be
\begin{align}
    \! [\mathbf{P}_{k,\BornRFS}^{\ell}]_{2,2} {=} [\mathbf{P}_{k,\BornRFS}^{\ell}]_{4,4} {=} [\mathbf{P}_{k,\BornRFS}^{\ell}]_{6,6} = \big( \tfrac{ \dot{r}^{\mathrm{max}} }{ 3 } \big)^2 = 
    1~\mathrm{m^2} \mathrm{s}^{-2}
    \! , \!
\end{align}
where $\dot{r}^{\mathrm{max}} {=} 3~\mathrm{m/s}$ is assumed maximum pedestrian speed (i.e., running) in 3D,
and the variances of width and height
\begin{align}
    [\mathbf{P}_{k,\BornRFS}^{\ell}]_{7,7} = \sigma_{\widthThreeDee}^2 \,, \qquad
    [\mathbf{P}_{k,\BornRFS}^{\ell}]_{8,8} = \sigma_{\heightThreeDee}^2 \,,
\end{align}
respectively, are based on the motion model.
Finally, equal weights for each GM component are used, and thus the birth spatial density~\eqref{eq:spatial-density-of-newly-born-objects} becomes
\begin{align}
    b_k(\mathbf{x}_k) = b(\mathbf{x}_k) = \textstyle \sum_{\ell = 1}^{10} \tfrac{1}{10} \cdot \mathcal{N}( \mathbf{x}_k;\ \mathbf{m}_{k,\BornRFS}^\ell, \, \mathbf{P}_{k,\BornRFS}^\ell ) \, . \label{eq:spatial-PDF:gaussian-mixture}
\end{align}
If MB birth is selected instead of the Poisson birth, the existence probabilities $r_{k,\BornRFS}^{\ell} {=} \tfrac{\beta}{ 10 }$, $\ell{=}1,\dots,10$ can be used.\\

The identified parameters capture certain characteristics of the given data, and they \emph{form} the entire SPO model.
In particular, the ground-truth BBs and the FRCNN detections from the \emph{entire} MOT-17 dataset (training subset) were used.
Under the assumption that the SPO model describes the given \emph{multi-object system} comprehensively, the corresponding model-based (i.e., the PMBM) filter provides the real posterior multi-object density, which is investigated in the next section.

\def\identity{i}
\section{Results and Discussion}\label{sec:results}
In this section, the model-based filter corresponding to the SPO model from Section~\ref{sec:SPO-Visual-tracking-model}, i.e., the approximate GM-PMBM filter further referred simply to as PMBM, is applied to the MOT-17 dataset (training subset) with inputs being \emph{exclusively} the FRCNN detections included in the dataset.
The set of ground-truth BBs at time step $k{=}1,\dots,K$ is denoted with $Y_k{=}\{ (\identity^1,\mathbf{y}_k^1), \dots, (\identity^{m_k},\mathbf{y}_k^{m_k})\}$, where $\identity^1,\dots,\identity^{m_k}$ are indices that denote identities of the BBs $\mathbf{y}_k^1,\dots,\mathbf{y}_k^{m_k}$.

\subsection{Implementation Details}
The Poisson initial multi-object density is taken to be the same as the birth density, i.e.,
\begin{align}
    p(X_0) = \e^{-\beta} \, \textstyle \prod_{\mathbf{x}_0 \in X_0} \, \beta \cdot b(\mathbf{x}_0).
\end{align}
At each time step, estimates are drawn using the sub-optimal Estimator 1~\cite[Sec.~VI.A]{PMBM:2018} for simplicity, with a \emph{non-informative} threshold of $0.5$ for existence probabilities.
Since only 2D ground-truth data are available, estimates are further projected into 2D using the unscented transform, see~\cite[Sec.~IV]{KrKoStDu:2024:IEEE}, which are denoted with $\mathbf{b}_k^{1},\dots,\mathbf{b}_k^{n_k}$ at time step $k$.
To connect the estimated BBs in time, i.e., to effectively output trajectories, the indices $\ell^1,\dots,\ell^{n_k}$ of the Bernoulli components from which the estimates are drawn from are saved along, yielding the set of estimates $B_k{=}\{ (\ell^1,\mathbf{b}_k^1), \dots, (\ell^{m_k},\mathbf{b}_k^{m_k})\}$.

Ellipsoidal gating was used with threshold~6.
To make room for computational constraints, the number of global hypotheses at each time step was capped at $25$, with a pruning threshold of $-100$ for log-weights, and Murty's \mbox{$M$-best} assignment algorithm was used with $M{=}3$.

\subsection{Performance Evaluation Process}
The CV scores such as \emph{multiple object tracking accuracy} (MOTA), \emph{higher order tracking accuracy} (HOTA), and \emph{identity F1} (IDF1) are often considered authoritative in the CV literature~\cite{HOTA:2021}. 
Since the CV scores may behave undesirably~\cite{Nguyen:Visual-metrics-trustworthy:2023,KrKoStXiSvFe:TGOSPA-params-visual:ArXiv}, the \emph{trajectory generalized sub-optimal assignment} (TGOSPA) metric is used, which is a mathematical metric on the space of sets of trajectories, and it allows for performance evaluation based on user preferences.
Note that the TGOSPA metric is approximated using the LP metric, see~\cite{TrajectoryGOSPA:2020}.

The TGOSPA metric can be decomposed into \emph{(i)} properly estimated objects, i.e., \emph{true positives} (TP), \emph{(ii)} missed objects, i.e., \emph{false negatives} (FN), \emph{(iii)} false objects, i.e., \emph{false positives} (FP), and \emph{(iv)} properly defined \emph{track switches} (Sw), see~\cite{TrajectoryGOSPA:2020}.
For simplicity, only the \emph{numbers} of objects forming \emph{(i)}-\emph{(iii)} denoted with $|$TP$|$, $|$FN$|$ and $|$FP$|$, respectively, are presented. 
In this paper, TGOSPA metric parameters are set based on~\cite{KrKoStXiSvFe:TGOSPA-params-visual:ArXiv} to \emph{(1)} allow reasonably distant estimates (e.g., predictive) to be considered as proper estimates and \emph{(2)} to consider reasonably long track changes as Sw.
In particular, the following parameters are used:
metric $d(\mathbf{b},\mathbf{y}) {=} 1 {-} \mathrm{IoU}(\mathbf{b},\mathbf{y}) $ with $\mathrm{IoU}$ being the \emph{intersection over union} (IoU) of the two bounding boxes $\mathbf{b}$ and $\mathbf{y}$;
cut-off $c {=} 0.5$ and exponent $p {=} 1.8$ to respect\footnote{
    The exponent $p$ is set so that an estimate $\mathbf{x}$ that is closer to a ground truth $\mathbf{y}$ than $a {=} \tfrac{c}{ \sqrt[p]{2} } {=} 0.34$ in the metric $d$ is considered \emph{better} compared to the case an estimate is not provided for $\mathbf{y}$ at all.
} preference \emph{(1)};
switching penalty $\gamma {=} c {\cdot} \sqrt[p]{ 10 } {=} 1.797$ so that track changes lasting for at least $10$ time steps are considered as switches thus respecting preference \emph{(2)}. 

Furthermore, we compute the cardinality mismatch, 
\begin{align}
    \text{cardinality mismatch} = \textstyle \big| \sum_{k=1}^K |B_k| - |Y_k| \big| \, , \label{eq:cardinality-mismatch}
\end{align}
i.e., the (absolute value of) the difference between the number of estimated BBs $\sum_{k=1}^K |B_k|$ and ground-truth BBs $\sum_{k=1}^K |Y_k|$.
Note that the cardinality mismatch can be assessed using the GOSPA metric~\cite{GOSPA:2017} with setting its parameter $\alpha{<}2$ small.

\def\thtb{($\uparrow$)} 
\def\tltb{($\downarrow$)} 
\definecolor{colFirst}{RGB}{0 255 0}
\def\a{25} 

\newcommand{\jrt}[2]{
    \begin{tabular}{|c|}
    \hline
        #1 \\
        #2 \\
    \hline
    \end{tabular}
}

\newcommand{\tbl}[1]{
    \begin{tabular}{c}
        #1 
    \end{tabular}
}

\begin{table*}[t] 
    \centering
    \caption{Results for the training subset of the MOT-17 dataset using FRCNN detections.}
    \label{tab:results}
    \definecolor{lightgray}{gray}{0.92}
    \rowcolors{3}{}{lightgray}
    \begin{tabular}{c|ccc|ccccc|c}
          & \multicolumn{3}{c|}{CV scores} & \multicolumn{5}{c|}{TGOSPA and its decomposition}  & \\[0.05cm]
    \jrt{PMBM}{\cellcolor{lightgray} SORT} & MOTA \thtb & HOTA \thtb & IDF1 \thtb & TGOSPA \tltb & $|$TP$|$ \thtb & $|$FN$|$ \tltb & $|$FP$|$ \tltb & Sw \tltb & \tbl{cardinality\\ mismatch} \hspace{-0.3cm}\tltb \\[0.3cm]
    & 31.936 & \bf 34.575 & \bf 39.378 & \multicolumn{1}{c|}{76.944} &\bf 6553 & \bf 12028 & 2507 & 32 & \bf 9521 \\
\cellcolor{white}\multirow{-2}{*}{MOT-17 02} & \bf 32.043 & 31.310 & 36.436 & \multicolumn{1}{c|}{\bf 74.204} & {5948} & 12633 & \bf 1649 & \bf 31 & 10984 \\\hline
    & \bf 55.746 & \bf 49.904 & \bf 57.647 & \multicolumn{1}{c|}{105.97} & \bf 26996 & \bf 20561 & 2181 & 64 & \bf 18380 \\
\cellcolor{white}\multirow{-2}{*}{MOT-17 04}& 53.998 & 49.797 & 56.416 & \multicolumn{1}{c|}{\bf 103.56} & 25877 & 21680 & \bf 2034 & \bf 63.5 & 19646 \\\hline
    & 38.716 & 40.964 & 50.205 & \multicolumn{1}{c|}{44.319} & \bf 3318 & \bf 3599 & 1009 & 20 & \bf 2590 \\
 \cellcolor{white}\multirow{-2}{*}{MOT-17 05} & \bf 45.121 & \bf 42.315 & \bf 53.534 & \multicolumn{1}{c|}{\bf 40.775} & 3021 & 3896 & \bf 310 & \bf 12.5 & 3586\\\hline
    & \bf 57.014 & \bf 47.826 & \bf 55.336 & \multicolumn{1}{c|}{30.947} & \bf 3145 & \bf 2180 & 273 & 19 & \bf 1907 \\
 \cellcolor{white}\multirow{-2}{*}{MOT-17 09} & 53.033 & 43.719 & 50.067 & \multicolumn{1}{c|}{\bf 30.929} & 2797 & 2528 & \bf 87 & \bf 17 & 2441 \\\hline
    & 36.109 & 36.287 & 40.562 & \multicolumn{1}{c|}{69.173} & \bf 7025 & \bf 5814 & 4282 & 46 & \bf 1532 \\
 \cellcolor{white}\multirow{-2}{*}{MOT-17 10} & \bf 49.552 & \bf 39.016 & \bf 43.807 & \multicolumn{1}{c|}{\bf 58.974} & 6519 & 6320 & \bf 1747 & \bf 42 & 4573 \\\hline
    & 52.014 & 45.284 & 48.155 & \multicolumn{1}{c|}{45.142} & \bf 5632 & \bf 3840 & 1138 & 25.5 & \bf 2666 \\
 \cellcolor{white}\multirow{-2}{*}{MOT-17 11} & \bf 55.468 & \bf 49.863 & \bf 53.937 & \multicolumn{1}{c|}{\bf 41.117} & 5255 & 4181 & \bf 308 & \bf 19.5 & 3873 \\\hline
    & 30.648 & 38.056 & 46.595 & \multicolumn{1}{c|}{63.690} & \bf 5306 & \bf 6336 & 2806 & \bf 26 & \bf 3530 \\
 \cellcolor{white}\multirow{-2}{*}{MOT-17 13} & \bf 45.834 & \bf 43.500 & \bf 50.337 & \multicolumn{1}{c|}{\bf 55.429} & 5515 & 6127 & \bf 1085 & 32 & 5042 \\
    \end{tabular}
    \vspace{-2mm}
\end{table*}

PMBM results are compared to SORT~\cite{SORT:2016} in Table~\ref{tab:results}.
It is important to note that the CV-based method SORT directly models BBs on the 2D image plane, uses M/N logic for track management, and determines detection-to-track assignments using $\mathrm{IoU}$; for more details see~\cite{SORT:2016}.

\subsection{Analysis of the Results}
From the readings of the CV scores, PMBM brings no significant improvement to SORT.
Considering the TGOSPA metric, PMBM is always worse than SORT.
However, PMBM estimates properly more ground-truth objects than SORT, i.e., the number of TPs is larger for PMBM for all videos.
As a result, the number of FNs is \emph{lower} for PMBM than for SORT.
While PMBM has more TPs, it also has more FAs than SORT.

Although the number of FPs should be as small as possible, it is to say that the sum ``$|$TP$|+|$FP$|$''${=} \sum_k |B_k|$ is closer to the number of ground-truth BBs $\sum_k |Y_k| {=}$``$|$TP$|+|$FN$|$'' for PMBM than for SORT.
That is, the cardinality mismatch\footnote{
    As a result, evaluating the results with the GOSPA metric with a small enough parameter $\alpha$, PMBM can yield considerable improvement compared to SORT in such a metric.
    Note that GOSPA is a metric on the space of finite sets of objects (i.e., BBs in the setting of this paper).
    Unlike TGOSPA, GOSPA is \emph{not} a metric on the space of sets of trajectories~\cite{TimeWeightedTGOSPA:2021}.
} is smaller for PMBM. 
Considering that the PMBM uses $L$ and $\eta$ in the model, see Section~\ref{sec:birth:expected-number}, the filter is prone to keeping the \emph{number} of estimates $|B_k|$ \emph{consistent} with $L {\cdot} \eta$.
Note that the FRCNN detector fails to detect a considerable portion of objects that were included to yield the estimated $L$ and $\eta$.
That is, to output better estimates of $|Y_k| $, the filter must have generated a considerable number of \emph{both} TPs and FPs.
\\

SORT overcomes the above problems by adopting logic-based heuristics.
The model-based approach premises, however, that when a model fits the reality, the model-based algorithm provides \emph{good} -- nearly optimal -- results.
If results are not \emph{good}, then either \emph{1)} the model does not fit the reality, \emph{2)} the estimator is problematic, or \emph{3)} the performance evaluation process fails to consider some key aspect or any combination of \emph{1)}--\emph{3)} hold.
For the above metrics, the evaluation process does not consider estimation error and consistency of the results, requiring further investigation~\cite{KrKoStXiSvFe:TGOSPA-params-visual:ArXiv,Coraluppi:Evaluation:2023,PobabilisticGOSPA:ArXiv:2024}.
The following discussion further shows that several components of the SPO model are problematic, causing the SPO model-data mismatch.

\subsection{Problematic Components of the Model}
During the SPO model parameters identification in Section~\ref{sec:SPO-Visual-tracking-model}, several simplifying assumptions were used, such as the Gaussian assumption or constant probabilities of detection and survival.
Such assumptions are common in the literature as they simplify the implementation of the PMBM filter.

For instance, the probability of detection $P_\ObjectGenRFS(\mathbf{x}_k)$ should clearly depend on whether the object $\mathbf{x}_k$ is in the tracking area or not, i.e., visible to the camera.
Although this could easily be adjusted, it can be argued that doing so would lead to practical complications leading to little improvement.
It turns out that the SPO model itself fails to consider the true nature of the probability of detection.

\subsubsection{Probability of Detection Dependent on Other Objects}

\begin{figure}
    \centering
    \includegraphics[width=0.45\textwidth,trim={1.5cm, 0.1cm, 3.6cm, 0.3cm},clip]{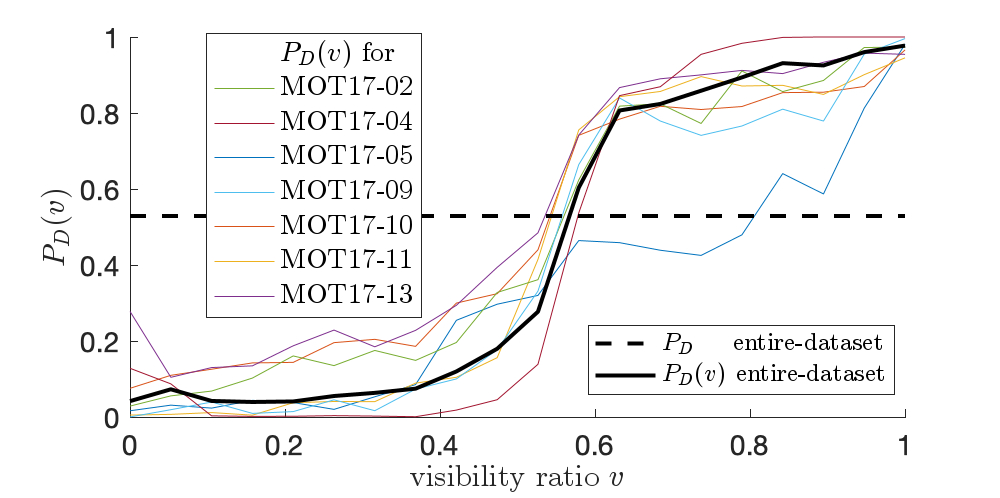}\\[-0.2cm]
    \caption{Estimated probability of detection $P_D(v)$ as a function of the visibility ratio $v{=}v(\mathbf{x},O)$ of an object $\mathbf{x}$ with respect to other objects $O$.
    The MOT-17 dataset was used, see footnote~\ref{footnote:estimation-tlwh-visibility}.
    }
    \label{fig:P_D(v):MOT-17}
    \vspace{-2mm}
\end{figure}

The estimation method~\cite{KrKoSt:2023_FUSION} could be easily adjusted to allow for drawing conditional estimation.
Taking the visibility ratio $v$ into account (that is available in the MOT-17 dataset), the estimated detection probability conditioned on $v$ is plotted in Fig.~\ref{fig:P_D(v):MOT-17} for the FRCNN detector.
Note that the visibility ratio $v {=} v(\mathbf{x},O)$ depends on the configuration of the other objects forming the set $O$.
Approximate algorithms considering that the probability of detection is dependent on other objects were discussed in e.g.,~\cite{Linh:VisualOccluLMB:2024}.

It can be seen from Fig.~\ref{fig:P_D(v):MOT-17} that the estimated function $P_\ObjectGenRFS(v)$ fits to all videos fairly.
While the same seems to hold for all measurement model parameters~\cite{KrKoSt:2023_FUSION}, SPO motion model parameters may be considerably different for each video.

\subsubsection{Scenario-dependent Parameters}
Besides having a different sampling period $T$, each video captures a different scene.
It is thus clear that different parameters should be valid for each video separately.
For instance, it may be beneficial to incorporate the acceleration vector into the state dynamics~\eqref{eq:3Dsystem:state} if objects may be standing still.
If objects can be expected to be running on the other hand (not the case in the MOT-17 dataset), the power spectral densities should be set larger, and the other single-object dynamical parameters should be adjusted appropriately as well.

\begin{table}
    \centering
    \caption{Estimated parameters of the SPO motion model.}
    \label{tab:estimated-parameters}
    \begin{tabular}{c|cc|cc||c}
         & \multicolumn{4}{c||}{values estimated from data} & \\[0.1cm]
         & $\Expect{ n_k }$ & $\var{ n_k }$ & $L$ & $\eta$ & $L \cdot \eta$ \\[0.13cm]
        MOT-17 02 & 30.968 & 14.281 & 9.990 & 2.000 & 19.980 \\
        MOT-17 04 & 45.292 & 10.680 & 19.010 & 1.171 & 22.373 \\
        MOT-17 05 & 8.264 & 4.599 & 3.715 & 2.124 & 7.891 \\
        MOT-17 09 & 10.143 & 4.375 & 6.827 & 1.143 & 7.802 \\
        MOT-17 10 & 19.632 & 10.497 & 7.508 & 1.743 & 8.108 \\
        MOT-17 11 & 10.484 & 4.822 & 4.194 & 1.933 & 8.108 \\
        MOT-17 13 & 15.523 & 73.831 & 4.234 & 2.933 & 12.418 \\[0.12cm]
        entire-dataset & 21.124 & 205.54 & 7.481 & 1.925 & 14.396 \\
        \multicolumn{1}{c}{} & \multicolumn{1}{c}{$\uparrow$} & \multicolumn{1}{c}{$\uparrow$} & & \multicolumn{1}{c}{} & \multicolumn{1}{c}{$\uparrow$} \\
        \multicolumn{6}{c}{\fbox{for the M/M/$\infty$ model, $\Expect{ n_k }$, $\var{ n_k }$, and $L \cdot \eta$ should be equal}}
    \end{tabular}
    \vspace{-2mm}
\end{table}

The mean life span $L$~\eqref{eq:L-estimates} and the parameter $\eta$~\eqref{eq:eta-estimates} of the Poisson process estimated for each video separately are shown in Table~\ref{tab:estimated-parameters}.
It can be seen that rather different parameters are valid for each video, especially regarding the parameter $L$.

Further analysis suggested that the Poisson process describes the arrival of newborn objects in time well.
According to the M/M/$\infty$ model, the number $n_k$ of objects is Poisson distributed with parameter $ \Expect{n_k} {=} L {\cdot} \eta$ for $k {\rightarrow} \infty$.
Assuming the system is captured already in the steady state, the product $L {\cdot} \eta$ should fit both $\Expect{n_k}$ and $\var{n_k}$ that can be estimated directly from the data.
It can be seen from Table~\ref{tab:estimated-parameters} that is not the case for most videos.
The data for each video individually appears to violate the assumption that the lifespan is exponentially distributed.
The analysis thus also suggests that a different lifespan model should be used.

\subsubsection{Further Problematic Components}
As already mentioned, camera motion and detection scores should be taken into account. 
Further, it is well-known that RFS-based methods can be used to incorporate measurements from multiple sources, offering the possibility to leverage, e.g., various visual features.

Assuming an object cannot disappear if it stands still, its survival should depend on its speed.
Since pedestrians are macroscopic objects, they cannot occupy the same place in space. 
Therefore, the birth spatial density should depend on the existing objects.
Moreover, objects may interact with each other, e.g., by following the same direction.
Finally, the clutter spatial density is likely to be different from uniform as the correlation of the size of a BB with its position clearly depends on the view captured by the camera, and clutter may further depend on the objects.

\section{Conclusion}\label{sec:conclusion}
In this paper, a fully model-based approach to visual tracking using 2D bounding boxes was adopted.
The standard-point object model commonly used in the radar tracking community was reviewed, and its parameters were identified for the visual tracking problem.
An approximate Gaussian mixture implementation of the corresponding model-based filter, which is the PMBM filter, was tested on the MOT-17 dataset, and its performance was evaluated using several methods.
The suitability of the model was assessed, elucidating particular components of the model causing a model-data mismatch.

\bibliographystyle{ieeetr}

\end{document}